\documentclass[12pt]{iopart}

\usepackage[dvips]{graphicx}
\usepackage{cite} 
\usepackage{amssymb}
\def\Hat#1{\hat{#1}}
\def\Dot#1{\dot{#1}}
\def\Ihathat{\mathbb{I}}
\def\Ahathat{\mathbb{A}}
\def\Dhathat{\mathbb{D}}
\def\Rhathat{\mathbb{R}}
\def\Xhathat{\mathbb{X}}
\def\Ghathat{\mathbb{G}}
\def\AA{\rme^{\Ahathat\cdot\tau}\rme^{\Ahathat'\cdot\tau'}}
\def\BB{\rme^{\Ahathat'\cdot\tau'}\rme^{\Ahathat\cdot\tau}}
\def\rhohat{\Hat{\rho}}
\def\journ#1{\textit{#1}}
\def\vol#1{\textbf{#1}}
\def\erf{\,\mathrm{erf}}

\begin{document}
\title[Influence of the atomic-wall collision elasticity on the CPT resonance shape]%
    {Influence of the atomic-wall collision elasticity on the coherent population trapping resonance shape}

\author{G A Kazakov$^{1}$, A N Litvinov$^{1}$, B G Matisov$^{1}$,
V I Romanenko$^{2}$, L P Yatsenko$^{2}$ and A V Romanenko$^{3}$}

\address{$^{1}$ St. Petersburg State Polytechnical University, 29, Polytechnicheskaya st, 
St. Petersburg 195251, Russia}
\ead{{kazjor@rambler.ru}}

\address{$^{2}$ Institute of Physics, Nat. Acad. Sci. of Ukraine, 46, Nauky Ave., 
Kyiv 03028, Ukraine}

\address{$^{3}$ Kyiv National Taras Shevchenko University, 4, Academician Glushkov Ave., 
Kyiv 03022, Ukraine}

%%%%%%%%%%%%%%%%%%%%%%%%%%%%%%%%%%%%%%%%%%%%%%%%%%%%%%%%%%%%%%%%%%%%%%%%%%%%%%%%%%%%%%%%%%%%%
%%%%%%%%%%%%%%%%%%%%%%%%%%%%%%%%%%%%%%%%%%%%%%%%%%%%%%%%%%%%%%%%%%%%%%%%%%%%%%%%%%%%%%%%%%%%%
\begin{abstract}
We studied theoretically {a} coherent population trapping resonance formation in
cylindrical cell {without buffer} {gas irradiated} by a narrow laser beam. We
take into account {non-zero probabilities} of elastic (``specular'') and
inelastic (``sticking'') collision between the atom and the cell wall. We {have
developed a} theoretical {model} based {on averaging} over the random Ramsey pulse
sequences of times that atom spent in and out of the beam. It is shown that
{the shape of coherent population trapping resonance line depends on the probability of elastic collision}.
\end{abstract}

\submitto{\jpb}

\maketitle                

%%%%%%%%%%%%%%%%%%%%%%%%%%%%%%%%%%%%%%%%%%%%%%%%%%%%%%%%%%%%%%%%%%%%%%%%%%%%%%%%%%%%%%%%%%%%%
%%%%%%%%%%%%%%%%%%%%%%%%%%%%%%%%%%%%%%%%%%%%%%%%%%%%%%%%%%%%%%%%%%%%%%%%%%%%%%%%%%%%%%%%%%%%%       
\section{Introduction}
\label{sect-1}

In the simplest case, the coherent population trapping (CPT) effect appears in three-level $\Lambda$-system, see \fref{fig-1}.
The atom is interacting with a two-frequency laser field coupling two metastable
states with {a} short-living state.
When the frequency difference between the laser field components is equal to the
frequency of the transition between {two} metastable states (two-photon resonance
conditions), the ``dark state'' {that} does not interact with {the} laser field appears.
This ``dark state'' is the coherent superposition of the metastable states. If
one scans the frequency difference of two laser components near the two-photon
resonance, one observes the CPT resonance, an abrupt decrease of the
fluorescence intensity~\cite{ref-1,ref-2,ref-3}.
CPT phenomenon allows to create a window in the absorption
spectrum~\cite{ref-4}.
This effect is used in compact frequency standard~\cite{ref-5,ref-6}, the
stimulated adiabatic passage technique of population transfer between the states
of atoms and molecules~\cite{ref-7}, laser cooling~\cite{ref-8} and magnetic field
measurements~\cite{ref-9,ref-10}.

It is well known~\cite{ref-11,ref-12,ref-13,ref-14,ref-15,ref-16,ref-17,ref-18,ref-19} that the CPT resonance shape is
affected by the movement of atoms between the beam zone illuminated by the laser
radiation and the dark zone.
The theory of the CPT resonance formation in wall-coated cells without buffer
gas based on averaging over random Ramsey sequences of times {which} the active atom
spent in the bright and dark zones was developed
in~\cite{ref-12,ref-13,ref-14}.
In these models, it was assumed that in \emph{any} atomic-wall collision the atom
sticks to the cell wall for some time.
During this time atom exchanges its kinetic energy with the cell wall and then
it is released back to the cell volume with alternated velocity that is not
correlated with atomic velocity before the collision.
Atoms make many transitions between these two zones during the lifetime of
metastable states coherence.
This cohrerence is assumed to be preserved in the collision~\cite{ref-20}.

Assumption about sticking collisions is based on the studies of relaxation of
polarization in Rubidium vapors in paraffin-coated cell performed by Bouchiat
and Brossel~\cite{ref-21}.
These studies showed that at least some atoms sticks to the cell wall for some
time in the wall collision act.
During this time, atom exchanges its kinetic energy with the cell wall and then
it is released back to the cell volume with alternated velocity.
At the same time, some part of atoms can collide elastically with the cell wall.

In this work, we make the generalization of the
model~\cite{ref-12,ref-13,ref-14} to the case when there is a {nonzero}
probability~$\alpha$ (reflection coefficient) of elastic collision between the
atom and the cell wall.
Examples of an elastic collision are evanescent wave mirrors~\cite{ref-22,ref-23} 
or magnetic gradient mirrors~\cite{ref-24,ref-25,ref-26} for cold
atoms. We assume that an elastic collision do not change velocity components
parallel to the cell wall (normal component changes sign).
We study the influence of the reflection coefficient~$\alpha$ on the CPT
resonance line shape at different intensities of laser radiation provided that
the width of the laser radiation is much smaller than the Doppler width of
{the} optical transition.

The density matrix equations describing the atomic polarization inside and
outside the laser beam are presented in section~\ref{sect-2}.
In section~\ref{sect-3}, we discuss the influence of elastic and inelastic
collisions on the atomic motion in the cell and derive the expression for the
excited state population.
The procedure of the ensemble averaging is described in section~\ref{sect-4}.
Results of calculations are presented in section~\ref{sect-5}.
Conclusions are made in section~\ref{sect-6}.

%%%%%%%%%%%%%%%%%%%%%%%%%%%%%%%%%%%%%%%%%%%%%%%%%%%%%%%%%%%%%%%%%%%%%%%%%%%%%%%%%%%%%%%%%%%%%
%%%%%%%%%%%%%%%%%%%%%%%%%%%%%%%%%%%%%%%%%%%%%%%%%%%%%%%%%%%%%%%%%%%%%%%%%%%%%%%%%%%%%%%%%%%%%
\section{Basic equations}
\label{sect-2}

\begin{figure}
\begin{center}
\includegraphics[scale=0.3]{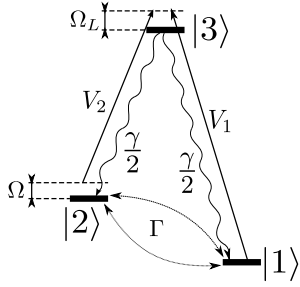}
\end{center}
\caption{$\Lambda$-scheme of the interaction 
of {the} three-level atom with the field, $V_{1}$ and~$V_{2} $ are the Rabi frequencies, 
$\Gamma $ and~$\gamma$ are relaxation rates for ground and excited states,  respectively, and
$\Omega _{L} $ and~$\Omega $ are {the} one-photon and two photon (Raman) detunings.}
\label{fig-1}
\end{figure}

Let us consider a gas of three-level atoms with the excited state~$|3\rangle$ 
and metastable lower states~$|1\rangle$ and~$|2\rangle$ 
(one of which can be the ground state). 
The field component with frequency~$\omega_{1}$ couples the states~$|1\rangle$ 
and~$|3\rangle$, whereas the component with frequency~$\omega_{2}$ 
couples the states~$|2\rangle$ and~$|3\rangle$ 
($\Lambda$-scheme of the atom-field interaction, see \fref{fig-1}). 
The interaction of {the} field with {the} atoms is described by Rabi frequencies 
$V_{1}=\frac{1}{2}\vec{d}_{13}\cdot\vec{E}_{1}/\hbar$, 
$V_{2}=\frac{1}{2}\vec{d}_{23}\cdot\vec{E}_{1}/\hbar$, respectively. 
Here~$\vec{d}_{13}$ and~$\vec{d}_{23}$ are the matrix elements of the atom's dipole momentum, 
$\vec{E}_{1}$ and~$\vec{E}_{2}$ are the amplitudes of the laser field components. 
The one-photon frequency detuning of the first component is $\Omega_{L} =\omega _{1}-\omega _{13}$,
whereas $\Omega=\omega _{1}-\omega _{2}-\omega_{12}$ is the two-photon detuning. 
Here, $\omega_{ij}$ is the transition frequency between~$i$ and~$j$ states. 
In our case $\omega_{12}$ belongs to the microwave range.

The equation for density matrix describing the atom in the field reads
\begin{equation}
\dot{\rho }_{ij} =-\frac{i}{\hbar } \sum _{k}\left[H_{ik} \, \rho _{kj} -\rho _{ik} H_{kj} \right]+\sum _{k,l}\Gamma _{ij,kl} \, \rho _{kl}\,, 
\label{eq-1}
\end{equation}
{where} $H_{ik}$ are the matrix elements of the Hamiltonian 
$\hat{H}=\hat{H}_{0}+\hbar\hat{V}(v_{z} ,t)$, 
$\hat{H}_{0}$ is the Hamiltonian of the free atom. 
The {operator of dipole interaction} of the atom with the laser field 
$\hbar\hat{V}(v_{z} ,t)$ depends on time~$t$ 
and on the component of atomic velocity $v_{z}$ along the field propagation direction; 
$\Gamma_{ij,kl}$ are the elements of the relaxation matrix. The nonzero matrix elements are
\begin{eqnarray}
\Gamma_{12,12}&=&\Gamma_{21,21}=-\Gamma,\nonumber\\
\Gamma_{11,22}&=&\Gamma_{22,11}=-\Gamma_{11,11}=-\Gamma_{22,22}=\Gamma,\nonumber\\
\Gamma_{13,13}&=&\Gamma_{31,31}=\Gamma_{23,23}=\Gamma_{32,32}=-\gamma',\\
\Gamma_{11,33}&=&\Gamma_{22,33}=\gamma/2,\nonumber\\
\Gamma_{33,33}&=&-\gamma.\nonumber
\label{eq-1a}
\end{eqnarray}
Here $\gamma'=(\gamma+\Gamma_{L}) /2$ is the relaxation rate for optical coherences {$\rho_{13}$, $\rho_{23}$}~\cite{ref-29}, 
$\Gamma_{L}$ is the spectral width of the laser radiation, 
$\gamma$ is the spontaneous relaxation rate for the excited state, 
$\Gamma$ is the relaxation rate {in} the ground state.

The light {absorption} in the cell is proportional to the excited state population~$\rho_{33}$. 
We consider the case of weak fields $V_{1,2}\ll \gamma$. 
In this case $\rho_{33}$ is much smaller than {the} populations 
$\rho_{11}$ and $\rho_{22}$ 
and can be expressed via $\rho_{11}$, $\rho_{22}$ and $\rho_{12}$ 
using standard adiabatic elimination procedure~\cite{ref-27}. 

Usually this procedure is applied in the case of large detuning 
$\Omega_{L}\gg{}V_{1,2}$, but it is also valid in the case of $\gamma\gg{}V_{1,2}$.
Let us consider for illustration an open $\Lambda$-system, when the atoms due to spontaneous emission decay 
from the excited state $|3\rangle$ to the states, different from $|1\rangle$ and $|2\rangle$.
It is well known that the irreversible loss can be incorporated in the Hamiltonian~\cite{Sho90}. 
For the open $\Lambda$-system in the rotating frame the Hamiltonian $\tilde{H}$ is
\begin{equation}
\tilde{H}=\hbar\left(\begin{array}{ccc}
0&0&V_{1}\\
0&-\Omega&V_{2}\\
V_{1}&V_{2}& -\Omega_{L}-\rmi{}\gamma/2
\end{array}\right).
\label{eq:Ham}
\end{equation}
The set of equations for the amplitudes is
\begin{eqnarray}
\rmi\dot{C}_{1}&=&V_{1}C_{3}, \nonumber \\
\rmi\dot{C}_{2}&=&-\Omega C_2+ V_{2}C_{3}, \label{eq:C2} \\
\rmi\dot{C}_{3}&=&V_{1}C_{1}+V_{2}C_{2} -\left(\Omega_{L}+\frac{\gamma}{2}\right)C_{3}. \nonumber
\end{eqnarray}
Typical evolution time of $C_1$ and $C_2$ is about $1/V_{1,2}$.
If $|\Omega_{L}-\rmi{}\gamma/2|\gg{}V_{1,2}$,
we can put $\dot{C}_{3}=0$ in~(\ref{eq:C2}) on the time scales
of evolution of $C_1$, $C_2$ and then express $C_3$ via $C_1$ and $C_2$, i.e.,
to perform the adiabatic elimination of the excited state.
In the case of closed $\Lambda$ system considered in our paper,
the Hamiltonian $H$ is the Hermitian part of $\tilde{H}$. Considerations
for the density matrix equations similar to ones from previous paragraph
can be performed analogically. Another important point is that in the case considered in this paper 
(room temperature and laser beam diameter of order of 1 mm or higher), the typical time of flight of the atom through the beam or dark zone
is much larger than $1/\gamma'$.

We also neglect the Doppler shift of the frequency of microwave transition or, what is the same, we suppose that the 
Doppler shift is equal for two optical transitions. This approximation is valid when the relative phase
between two optical fields doesn't changes significantly on the length of typical path of the atom (Dicke narrowing~\cite{ref-28}).
In~\cite{kaz07, lit09}, it was shown that for the longitudinal cell size smaller than $\lambda_{21}/4$ 
(where $\lambda_{21}=2 \pi c /\omega_{21}$ is the wavelength of transition between the states $|1\rangle$ and $|2\rangle$), 
the shape of CPT resonance in coated cell without buffer gas practically does not depend on the cell length. Therefore 
we can neglect the Doppler broadening of the microwave transition if the cell size is less than $\lambda_{21}/4$ that  is
equal to 1.1~cm for microwave transition in $^{87}$Rb atom.

Using the normalization condition
\begin{equation}
\rho_{11}+\rho_{22}=1
\label{eq-2}
\end{equation}
and denoting $\hat\rho=\{f,R,J\}$, where $f=\rho _{11} -\rho _{22}$, 
$R=\mathop{\mathrm{Re}}(\rho _{12})$, $J=\mathop{\mathrm{Im}}(\rho _{12})$, 
we can rewrite the equations for the density matrix for the ensemble of atoms in the laser beam in the form:
\begin{eqnarray}
\dot{f}&=&G\frac{V_{2}^{2}-V_{1}^{2}}{\gamma'}-(W+\Gamma)f-4F\frac{V_{1}V_{2}}{\gamma'} J\,,\nonumber \\ 
\dot{R}&=&-G\frac{V_{1}V_{2}}{\gamma'} -(W+\Gamma)R-(\Omega-\Delta)J\,,\label{eq-3} \\
\dot{J}&=&F\frac{V_{1}V_{2}}{\gamma'}\,f +(\Omega-\Delta)R-(W+\Gamma)J \,.\nonumber
\end{eqnarray}
Here $G=G(v_{z})$ and $F=F(v_{z})$  denote the real and the imaginary parts of the expression 
$\gamma'/\bigl(\gamma'-i(\Omega _{L} -kv_{z})\bigr)$ respectively, 
$k=\omega_{1}/c$ is the wave number of the optical radiation,
$\Delta=F(V_{1}^{2}-V_{2}^{2})/\gamma'$ 
is the light shift, $W=G(V_{1}^{2}+V_{2}^{2} )/\gamma'$ denotes the {optical pumping rate}. 
The excited state population expressed in terms of $f$ and $R$ is
\begin{equation}
\rho_{33}=\frac{W}{\gamma}+\frac{G}{\gamma\gamma'}\bigg[(V_{1}^{2} -V_{2}^{2})\cdot f
+4V_{1}V_{2}\cdot R\bigg]\,.
\label{eq-4}
\end{equation}

To simplify the calculation procedure we use the matrix notation formalism similar to~\cite{ref-30}.
The set of equations~(\ref{eq-3}) can be symbolically presented as
\begin{equation}
\Dot{\Hat{\rho}}(v_{z},t)
=\Ahathat(v_{z})\Hat{\rho}(v_{z},t)+\Hat{B}(v_{z})\,,
\label{eq-5}
\end{equation}
Also we can write the expression~(\ref{eq-4}) in the form
\begin{equation}
\rho_{33} (v_{z} ,t)=\Hat{U}^{T} (v_{z} )\Hat{\rho }(v_{z} ,t)+V(v_{z})\,.
\label{eq-6}
\end{equation}
Furthermore, {for sake of brevity} we will skip the argument $v_{z}$ 
in $\Ahathat$, $\Hat{B}$, $\Hat{U}$ and~$V$. 

The evolution equation for the density matrix of atoms outside the laser beam can be obtained 
from equations~(\ref{eq-3}) by substitution $V_{1}=V_{2}=W=\Delta=0$. 
One can rewrite the resulting set of equations symbolically as
\begin{equation}
{\Dot{\Hat{\rho}}(v_{z},t)=\Ahathat'\Hat{\rho}(v_{z},t)} .
\label{eq-7}
\end{equation}
Note that $\Ahathat'$ does not depend on~$v_{z}$.

Finally, the density matrix of the atom is described by the equation~(\ref{eq-5}) 
inside the laser beam and by the equation~(\ref{eq-7}) outside the beam. 
The solution of the equation~(\ref{eq-5}) can be presented as
\begin{equation}
\rhohat(t)
=\Bigl(\Ihathat-\rme^{\Ahathat\cdot(t-t_{0})}\Bigr)\rhohat_{S} 
+\rme^{\Ahathat\cdot(t-t_{0})}
\rhohat(t_{0})\,,
\label{eq-8}
\end{equation}
where $\Hat{\rho }_{S} =-\Ahathat^{-1}\Hat{B}$ is {the} stationary solution 
of~(\ref{eq-5}) and $t_{0}$ is the initial {time} for which the density matrix is known,
{$\Ihathat$ is the identity matrix}. 
The solution of equation~(\ref{eq-7}) is written in a similar way:
\begin{equation}
\rhohat(t)=\rme^{\Ahathat'\cdot(t-t_{0})}\rhohat(t_{0})\,.
\label{eq-9}
\end{equation}

%%%%%%%%%%%%%%%%%%%%%%%%%%%%%%%%%%%%%%%%%%%%%%%%%%%%%%%%%%%%%%%%%%%%%%%%%%%%%%%%%%%%%%%%%%%%%
%%%%%%%%%%%%%%%%%%%%%%%%%%%%%%%%%%%%%%%%%%%%%%%%%%%%%%%%%%%%%%%%%%%%%%%%%%%%%%%%%%%%%%%%%%%%%
\section{Atomic motion in the cell}
\label{sect-3}

\begin{figure}
\begin{center}
\includegraphics[scale=0.2]{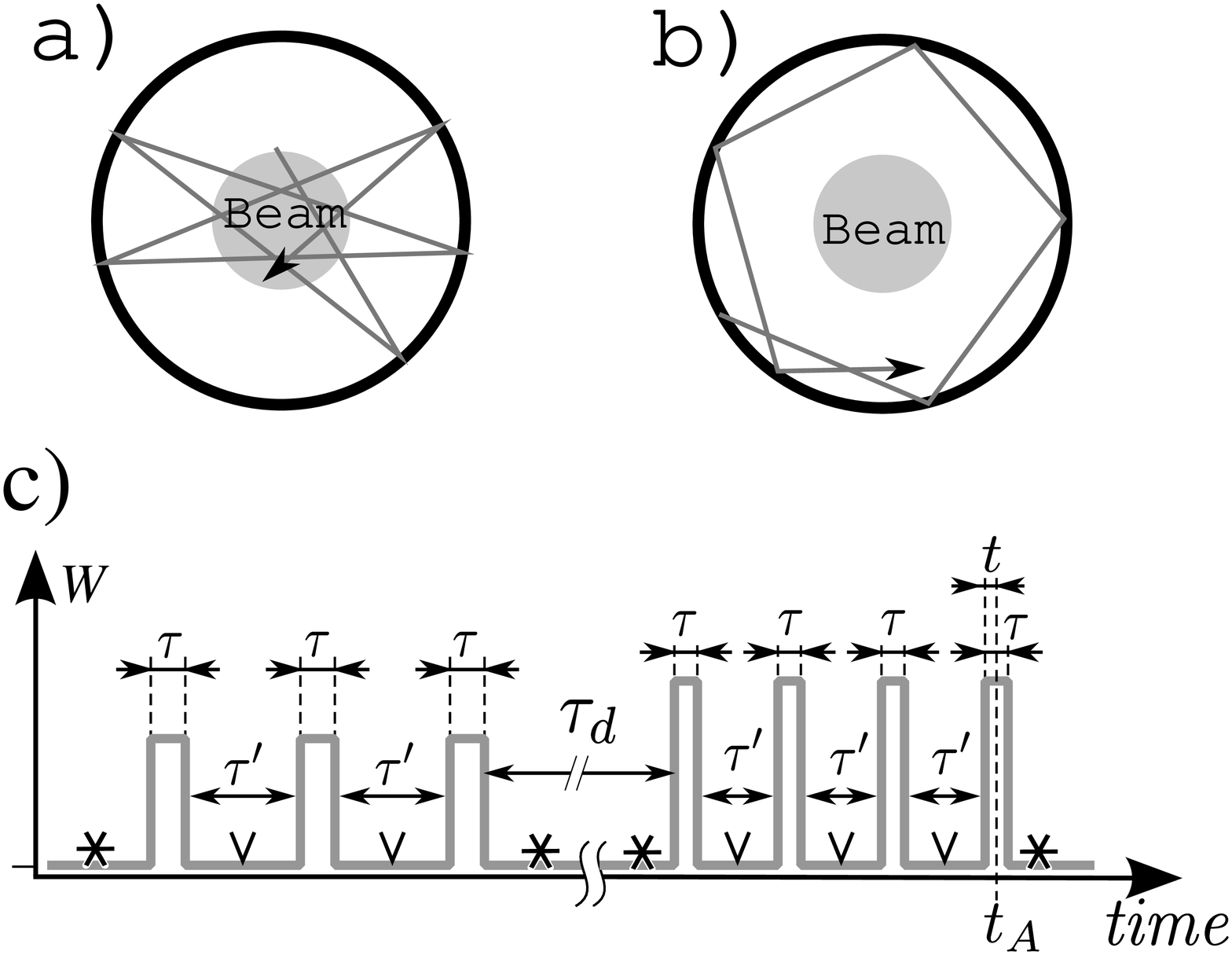}
\end{center}

\caption{The trajectory of a single atom in a cell in {the beam passing regime} (a)
and in the dark regime (b). (c) dependence of the optical pumping 
rate~$W$ on time. The magnitude of~$W$ depends on the velocity of the
atom that causes the Doppler shift of the atom's transition frequencies.    The
checkmarks show the {times of elastic} atom-wall collisions and the asterisks
show the {times of inelastic} collisions.}
\label{fig-2}
\end{figure}

In this paper, we consider a cylindric cell illuminated by a co-axial cylindric
laser beam. Let us imagine first that the atom undergoes only elastic
collisions with the cell walls. Therefore there are only two possibilities:
either the atom passes through the beam zone during each pass through the cell
as {it is} shown in \fref{fig-2}(a), or it does not enter the beam zone at all
(\fref{fig-2}(b)). In the first case we say that the atom is in the \emph{beam passing regime}, 
while in the second case the atom is in the \emph{dark regime}. 
Obviously, the resonance is formed only by the atoms in the
{beam passing regime}. Let us consider {some of such atoms}. At the
observation time~$t_{A} $ this atom was in the beam zone during some time~$t$
continiously. Before the enter to the beam zone the atom was in the dark zone
during time $\tau'$. Before this the atom was in the beam during time~$\tau$.
In turn, before this the atom was in the dark zone during time $\tau'$ etc.
Using equation~(\ref{eq-8}) and~(\ref{eq-9}), we obtain        
\begin{eqnarray}
\Hat{\rho}(t_{A})&=&
\Bigl(\Ihathat-\rme^{\Ahathat\cdot t}\Bigr)\rhohat_{S}
+\rme^{\Ahathat\cdot t}\rme^{\Ahathat'\cdot\tau'}
\Bigl[
\Bigl(\Ihathat-\rme^{\Ahathat\cdot \tau}\Bigr)\rhohat_{S}\nonumber\\
&&+\AA 
\Bigl[
\Bigl(\Ihathat-\rme^{\Ahathat\cdot \tau}\Bigr)\rhohat_{S}+\AA
\Bigl[\dots\Bigr]
\Bigr]
\Bigr] \nonumber\\
&=&\biggl[
\Bigl(\Ihathat-\rme^{\Ahathat\cdot t}\Bigr)
+\rme^{\Ahathat\cdot t}\rme^{\Ahathat'\cdot\tau'}
\Bigl(
\Ihathat-\AA
\Bigr)^{-1}\Bigl(\Ihathat-\rme^{\Ahathat\cdot \tau}\Bigr)
\biggr]\rhohat_{S} .
\label{eq-10}
\end{eqnarray}

%%%%%%%%%%%%%%%%%%%%%%%%%%%%%%%%%%%%%%%%%%%%%%%%%%%%%

%%%%%%%%%%%%%%%%%%%%%%%%%%%%%%%%%%%%%%%%%%%%%%%%%%%%%\sigma

Now let us consider more common case when some {of} the wall collisions change atomic
velocity. After such a collision the atom being initially in the dark regime
can either stay in this {regime} (which does not affect the equations for the
density matrix) or switch to the {beam passing regime}. An atom that was being in
the beam passing regime before the collision can either pass into the dark
regime or get back to the beam passing regime, but with different value of the
velocity and angle of reflection with the wall of the cell (and therefore with
different values {of}~$\tau$ and $\tau'$). {For sake} of definiteness, let us call
the time of entry of the atom into the beam zone after an inelastic collision as
the beginning of the beam passing regime, and the time of exit from the
beam zone as the end, respectively. Obviously, if the atom undergoes~$N$ elastic
collisions with the cell wall in the beam passing regime, then it goes~$N$ times
through the dark zone and $N+1$ times through the beam zone (see \fref{fig-2}(c)).

The density matrix $\rhohat_{e}$ describing the atom at the time of exit from
the beam passing regime can be obtained similarly {to} the expression~(\ref{eq-10}).
It reads                          
\begin{eqnarray}
\rhohat_{e}&=&
\biggl[\Ihathat+\AA+\dots+\Bigl(\AA\Bigr)^{N}\biggr]
\Bigl(\Ihathat-\rme^{\Ahathat\cdot\tau}\Bigr)\rhohat_{S}  \nonumber\\
&+&\Bigl(\AA\Bigr)^{N}\rme^{\Ahathat\cdot\tau}\rhohat_{b} \nonumber\\
&=&\Bigl(\Ihathat-\AA\Bigr)^{-1}
\biggl[\Ihathat-\Bigl(\AA\Bigr)^{N+1}\biggr]
\Bigl(\Ihathat-\rme^{\Ahathat\cdot\tau}\Bigr)\rhohat_{S} \nonumber\\
&+&\Bigl(\AA\Bigr)^{N}\rme^{\Ahathat\cdot\tau}\rhohat_{b}\,.
\label{eq-11}
\end{eqnarray}
Here, $\rhohat_{b}$ is the density matrix of the atom at the
time of beginning of the beam passing regime. It can be expressed via the
density matrix $\rhohat_{e,-1}$ for the previous time of exit
from the beam passing regime (it is indicated by the index ``$-1$'') using the
expression~(\ref{eq-9}) 
\begin{equation}
\rhohat_{b} =\rme^{\Ahathat'\cdot\tau_{d}}\rhohat_{e,-1}\,,
\label{eq-12}
\end{equation}   
where $\tau_{d}$ is the time {that} atom spent in the dark. Similar to~(\ref{eq-11}), 
one can obtain the expression for the density matrix at the observation time $t_{A}$:
\begin{eqnarray}
\rhohat(t_{A})&=&\rme^{\Ahathat\cdot t}\rme^{\Ahathat'\cdot \tau'}\Bigl(\Ihathat-\AA\Bigr)^{-1}
\biggl[\Ihathat-\Bigl(\AA\Bigr)^{n}\biggr]
\Bigl(\Ihathat-\rme^{\Ahathat\cdot\tau}\Bigr)\rhohat_{S}
\nonumber \\
&&+\Bigl(\Ihathat-\rme^{\Ahathat\cdot t}\Bigr)\rhohat_{S}
+\rme^{\Ahathat\cdot t}\Bigl(\BB\Bigr)^{n}\rhohat_{b}
\,.
\label{eq-13}
\end{eqnarray}

%
%%%
The expression for the excited state population $\rho_{33}$ reads
\begin{eqnarray}
\rho_{33}&=&\Hat{U}^{T}
\biggl\{
\rme^{\Ahathat\cdot t}\rme^{\Ahathat'\cdot \tau'}
\Bigl(\Ihathat-\AA\Bigr)^{-1}\biggl[\Ihathat-\Bigl(\AA\Bigr)^{n}\biggr]\Bigl(\Ihathat-\rme^{\Ahathat\cdot\tau}\Bigr)\rhohat_{S}
\nonumber \\
&&+\Bigl(\Ihathat-\rme^{\Ahathat\cdot t}\Bigr)\rhohat_{S}
+\rme^{\Ahathat\cdot t}\Bigl(\BB\Bigr)^{n}\rhohat_{b}
\biggr\}+V\,.
\label{eq-14}
\end{eqnarray}

%
%%%
Here, $n$ is the total number of collisions in the beam passing regime
immediately prior to observation time $t_{A}$. Averaging the expressions~(\ref{eq-14}), 
(\ref{eq-11}) and~(\ref{eq-12}) over all the
atoms in the beam zone, we obtain the average excited state population 
$\langle\rho _{33}\rangle $ that determines the absorption of light in the cell.

%%%%%%%%%%%%%%%%%%%%%%%%%%%%%%%%%%%%%%%%%%%%%%%%%%%%%%%%%%%%%%%%%%%%%%%%%%%%%%%%%%%%%%%%%%%%%
%%%%%%%%%%%%%%%%%%%%%%%%%%%%%%%%%%%%%%%%%%%%%%%%%%%%%%%%%%%%%%%%%%%%%%%%%%%%%%%%%%%%%%%%%%%%%
\section{Ensemble averaging}
\label{sect-4}

%%%%%%%%%%%%%%%%%%%%%%%%%%%%%%%%%%%%%%%%%%%%%%%%%%%%%%%%%%%%%%%%%%%%%%%%%%%%%%%%%%%%%%%%%%%%%
\subsection{General scheme of averaging}

Let us consider in detail the averaging of~(\ref{eq-14}). In this paper,
we consider the excitation of the CPT-resonance by a laser whose spectral width
is much smaller than the Doppler width of the optical transition. Hence, the
main contribution to the resonance is given by the atoms with small longitudinal
velocity, that are rarely colliding with the ends of the cell. We assume that
the collisions of the atoms in the beam passing regime with the side walls of
the cell do not change~$v_{z}$. The average population~$\langle\rho _{33}\rangle $ is
\begin{eqnarray}
\langle\rho_{33}\rangle&=&
\Bigl\langle\Hat{U}^{T}\Bigl(\Ihathat-\rme^{\Ahathat\cdot t}\Bigr)\rhohat_{S}\Bigr\rangle
+\Bigl\langle\Hat{U}^{T}\rme^{\Ahathat\cdot t}\Bigl(\BB\Bigr)^{n}\Bigr\rangle\left\langle\rhohat_{b}\right\rangle
\nonumber \\
&&+\langle V\rangle+\biggl\langle\Hat{U}^{T}\rme^{\Ahathat\cdot t}\rme^{\Ahathat'\cdot \tau'}\Bigl(\Ihathat-\AA\Bigr)^{-1}\nonumber\\
&&\times\biggl[\Ihathat-\Bigl(\AA\Bigr)^{n}\biggr]
\Bigl(\Ihathat-\rme^{\Ahathat\cdot\tau}\Bigr)\rhohat_{S}\biggr\rangle,
\label{eq-15}
\end{eqnarray}
where the angle brackets denote the averaging over the $z$-component of
the velocity~$v_{z}$, over the times $t$, $\tau$, $\tau'$, $\tau_{d}$ and
over the numbers~$n$ and~$N$ of elastic collisions in the beam passing regime
prior the observation time~$t_{A}$ and prior the exit from the beam passing
regime (in the expression for $\langle\rhohat_{b}\rangle$) 
respectively. The averaged density matrix $\langle\rhohat_{b}\rangle$ 
of the atom entering the beam passing regime can be obtained by averaging of expressions~(\ref{eq-11}) 
and~(\ref{eq-12}):
\begin{eqnarray}
\langle\rhohat_{e}\rangle=\Hat{C}+\Dhathat\langle\rhohat_{b}\rangle  ,
\label{eq-16}
\\
\langle\rhohat_{b}\rangle=\langle \rme^{\Ahathat'\cdot\tau_{d}}\rangle\langle\rhohat_{e}\rangle
=\langle \rme^{\Ahathat'\cdot\tau_{d}}\rangle  
\Bigl(
\Hat{C}+\Dhathat\langle\rhohat_{b}\rangle
\Bigr);
\label{eq-17}
\end{eqnarray}
hence it is easy to find 
\begin{equation}
\langle\rhohat_{b}\rangle=
\Bigl(\Ihathat-\langle \rme^{\Ahathat'\cdot\tau_{d}}\rangle \Dhathat
\Bigr)^{-1}\langle \rme^{\Ahathat'\cdot\tau_{d}}\rangle\Hat{C} .
\label{eq-18}                                                                                              
\end{equation}
Here
\begin{eqnarray}
\Hat{C}=
\biggl\langle
\Bigl(\Ihathat-\AA\Bigr)^{-1}
\biggl[
\Ihathat-\Bigl(\AA\Bigr)^{N+1}
\biggr]
\Bigl(\Ihathat-\rme^{\Ahathat\cdot\tau}\Bigr)\rhohat_{S}
\biggr\rangle ,
\label{eq-19}\\
\Dhathat
=\biggl\langle
\Bigl(\AA\Bigr)^{N}\rme^{\Ahathat\cdot\tau}
\biggr\rangle\,.
\label{eq-20}
\end{eqnarray}
Let us average over~$n$ and~$N$ first. Assuming that the probability of the
elastic scattering~$\alpha$ does not depend on the velocity of an atom and its
internal state, we obtain the probability~$P(n)=(1-\alpha )\alpha ^{n}$ for~$n$
consecutive elastic collisions. Hence, for any matrix $\Rhathat$ we get 
the matrix $\Rhathat^n$ averaged over $n$ as
\begin{equation}
\bigl\langle\Rhathat^n\bigr\rangle_{n}=\sum\limits_{n=0}^{\infty}P(n)\Rhathat^{n}
=(1-\alpha)\bigl(\Ihathat-\alpha\Rhathat\bigr)^{-1}\,.
\label{eq-21}
\end{equation}
Exactly the same kind of distribution~$P(N)=(1-\alpha )\alpha ^{N}$ is
valid for a total number of elastic collisions $N$ in the beam passing regime; therefore, the matrix
$\Rhathat^N$ averaged over $N$ is described by the same expression: 
$\bigl\langle\Rhathat^N\bigr\rangle_{N}=\bigl\langle\Rhathat^n\bigr\rangle_{n}$.

The next steps are {the} averaging over times $t$, $\tau$, $\tau'$ and $\tau _{d}$.
It is obvious that the averaging over the {time $\tau_{d}$} of atomic stay in the dark regime
is independent on $t$, $\tau$, $\tau'$, while $\tau$ and $\tau'$, 
in general, are determined by transversal component of the atomic velocity
in {the} beam passing regime and by the angle between the velocity and the wall of the
cell. The time $t$ is determined also by atom's position in {the} beam zone at the
observation time~$t_{A} $. Strictly speaking, one must integrate over four
variables $t$, $\tau$, $\tau'$ and $v_{z}$. However, we make a simplifying
assumption which reduces the number of integration variables to 1.

In the beam passing regime atom goes~$N$ times through the dark zone and~$N+1$
times through the beam zone. Each passage through the dark zone takes time~$\tau'$ 
and each passage through the beam zone takes time~$\tau$. Obviously, for each
passage the times $\tau'$ and $\tau$ are not independent. According to our
assumption we consider times of \emph{different} passages of the dark and
light zones as independent random variables. We emphasize that this assumption
refers only to ensemble averaging. In this case the averaged values of
$\rme^{\Ahathat\cdot \tau } $,
$\rme^{\Ahathat\cdot T} $,
$\rme^{\Ahathat'\cdot \tau '} $ and
$\rme^{\Ahathat'\cdot \tau_{d}} $ in~(\ref{eq-15})--(\ref{eq-20}) 
can be calculated independently.
Consider, for example, the averaged expression of
$\rme^{\Ahathat\cdot \tau } $. Its easy to present
this expression in the form~\cite{ref-12}
\begin{equation}
\bigl\langle \rme^{\Ahathat\cdot\tau}\bigr\rangle_{\tau}
=
\Xhathat
\left(
\begin{array}{ccc}
\bigl\langle \rme^{\lambda_{1}\tau}\bigr\rangle_{\tau} & 0 & 0   \\
0 & \bigl\langle \rme^{\lambda_{2}\tau}\bigr\rangle_{\tau} & 0 \\
0 & 0 & \bigl\langle \rme^{\lambda_{3}\tau}\bigr\rangle_{\tau}
\end{array}
\right) 
\Xhathat^{-1},
\label{eq-22}
\end{equation}
where $\Xhathat=(\Hat{X}^{(1)},\Hat{X}^{(2)},\Hat{X}^{(3)})$ 
is the matrix constructed {of} the eigenvectors of $\Ahathat$ corresponding to the eigenvalues 
$\lambda_{1},\lambda _{2},\lambda _{3}$. 
Therefore, averaging over~$\tau$ is reduced to {the calculation of integrals}
\begin{equation}
\langle \rme^{\lambda \tau }\rangle_{\tau} =g_{\tau}(\lambda)=\int_{0}^{\infty}\rme^{\lambda\tau}\cdot f_{\tau}(\tau)\, \rmd\tau \,,
\label{eq-23}
\end{equation}
that can be performed analytically. Here, $f_{\tau}$ is the probability density
function of the random variable~$\tau$. Averaging over $t$, $\tau'$ and
$\tau_{d} $ is analogous to~(\ref{eq-23}) with functions $g_{\tau'}(\lambda )$, 
$g_{t}(\lambda)$ and $g_{\tau_{d}}(\lambda)$. Note that $\mathop{\mathrm{Re}}(\lambda)<0$.

%%%%%%%%%%%%%%%%%%%%%%%%%%%%%%%%%%%%%%%%%%%%%%%%%%%%%%%%%%%%%%%%%%%%%%%%%%%%%%%%%%%%%%%%%%%%%
\subsection{Averaging over $\tau$ and t}

First of all, we find the probability density functions~$f_{\tau}$ and~$f_{t}$
of the random variables~$\tau$ and~$t$. The time~$\tau$ that the atom spend in the
beam zone is
\begin{equation}
\tau =\frac{2\sqrt{r^{2} -R^{2}\sin^{2}\varphi}}{v_{\perp}}\,,
\label{eq-24}
\end{equation}
where $\varphi$ is the angle between the atomic velocity projection on the plane orthogonal 
to the cell axis and the normal vector to the cell surface, 
$R$ is the radius of the cell, $r$ is the radius of the beam and~$v_{\perp}$
is the component of the velocity orthogonal to~$z$. 
Probability density function of the transversal velocity~$v_{\perp}$ 
is the two-dimensional Maxwell distribution 
$M_{2}(v_{\perp})=2\exp(-v_{\perp}^{2}/v_{T}^{2})v_{\perp}/v_{T}^{2}$ 
where $v_{T}=\sqrt{2k_{B} T/m}$ is the most probable atomic velocity, 
$k_{B}$ is {the} Boltzmann constant, $T$ is the temperature, 
$m$ is the mass of the atom. 
The probability density $\Phi(\varphi)$ of the random variable~$\varphi$ 
is proportional to $\cos\varphi$~\cite{ref-14,ref-31,ref-32}. 
Taking into account that in the beam passing regime the 
$\varphi$ takes values between $0$ and $\arcsin(r/R)$, 
we obtain $\Phi(\varphi )=R\cos\varphi/r$. 
Then, the cumulative distribution function $F_{\tau}(\tau)$ of the random variable ~$\tau$ is
\begin{eqnarray}
F_{\tau } (\tau )&=&\int\limits_{0}^{\arcsin(r/R)}\!\!\!\!\!\!\!\!
\Phi(\varphi )\!\!\!\!\!\!\!\!\int\limits_{\frac{2\sqrt{r^{2} -R^{2}\sin ^{2}\varphi}}{\tau}}^{\infty}\!\!\!\!\!\!\!\!{}M_{2}(v_{\perp})\,\rmd{}v_{\perp}\,\rmd\varphi
\nonumber \\
&=&\frac{\tau v_{T} \sqrt{\pi}}{4\rmi{}r}\,\rme^{-\frac{4r^{2}}{\tau^{2}v_{T}^{2}}}\erf\left(\frac{2\rmi{}r}{\tau v_{T}}\right)\,,
\label{eq-25}
\end{eqnarray}
and the probability density function is
\begin{equation}
f_{\tau} (\tau)=\frac{\rmd{}F_{\tau}(\tau)}{\rmd\tau} 
=-\frac{1}{\tau}+\frac{v_{T}\sqrt{\pi}}{4\rmi{}r}\, \rme^{-\frac{4r^{2}}{\tau^{2}v_{T}^{2}}}
\erf\left(\frac{2\rmi{}r}{\tau v_{T} }\right)
\left[1+\frac{8r^{2}}{\tau^{2}v_{T}^{2} }\right]\,.
\label{eq-26}
\end{equation}
Note that the average time that the atom moves through the beam zone is 
\begin{equation}
\bar{\tau}=\int_{0}^{\infty}f(\tau)\tau \,\rmd\tau =\int _{0}^{\infty }(1-F(\tau))\,\rmd\tau  
=\frac{\pi^{3/2}}{2}\frac{r}{v_{T}}\,.
\label{eq-27}
\end{equation}

Let us find the probability distribution function~$f_{t}$ of time~$t$ that atom
stays continuously in the beam zone before the observation time~$t_{A} $. The
averaging is carried out over all the atoms in the beam zone. Let $\rmd{}n/\rmd{}t$ atoms arrive
to the beam zone per unit of time. Then the total number
of atoms in the beam zone is
\begin{equation}
N_{tot}=\frac{\rmd{}n}{\rmd{}t}\int_{0}^{\infty }(1-F_{\tau }(t'))\,\rmd{}t' .
\label{eq-28}
\end{equation}
Here, we took into account that~$1-F_{\tau}(t')$ is the atom's probability to
remain in the beam zone after time~$t'$ from the entrance to the beam.
Therefore the number of atoms that have already spent time~$t$ or more in {the} beam
zone is
\begin{equation}
N_{t+}=\frac{\rmd{}n}{\rmd{}t}\int_{t}^{\infty }(1-F_{\tau}(t'))\,\rmd{}t'.
\label{eq-28a}
\end{equation}
Hence, the cumulative distribution function $F_{t}(t)$ of the random variable~$t$ reads
{\begin{equation}
F_{t}(t)=1-\frac{\int_{t}^{\infty }\bigl(1-F_{\tau }(t')\bigr)\,\rmd{}t'}{\int_{0}^{\infty }\bigl(1-F_{\tau }(t')\bigr)
\,\rmd{}t'}\,,
\label{eq-29}
\end{equation}
and the corresponding probability density function}
\begin{equation}
f_{t}(t)=\frac{1-F_{\tau}(t)}{\int _{0}^{\infty }\bigl(1-F_{\tau }(t')\bigr)\,\rmd{}t'} 
=\frac{1-F_{\tau }(t)}{\bar{\tau}}\,,
\label{eq-30}
\end{equation}
is
\begin{equation}
f_{t}(t)=\frac{2v_{T}}{\pi^{3/2}r} 
\left[
1-\frac{\tau v_{T} \sqrt{\pi}}{4\rmi{}r}\rme^{-\frac{4r^{2}}{\tau ^{2}v_{T}^{2}}}
\erf\left(\frac{2\rmi{}r}{\tau v_{T} }\right)
\right]\,.
\label{eq-31}
\end{equation}

\begin{figure}
\begin{center}
\includegraphics[scale=1]{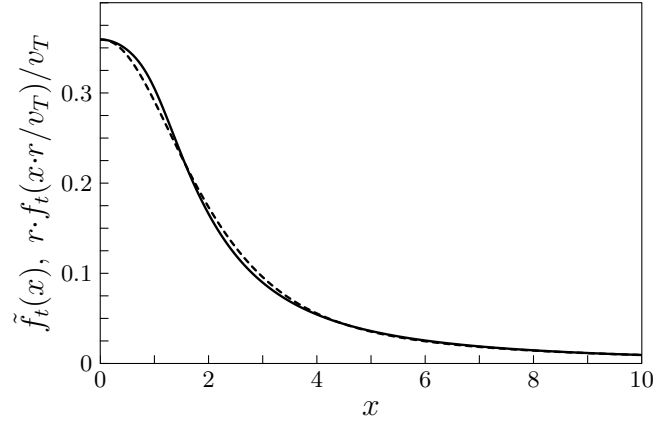}
\end{center}
\caption{Plots of $\tilde{f}_{t} \left(x\right)$ (dashed line) and $r{\cdot}f_{t} (x{\cdot}r/v_{T} )/v_{T} $ (solid line)}
\label{fig-3}
\end{figure}

Using~(\ref{eq-30}) it is easy to express $g_{\tau } (\lambda )$ in terms of $g_{t} (\lambda )$:
\begin{equation}
g_{\tau }(\lambda)=1+\lambda\,\bar{\tau }\,g_{t}(\lambda )
\label{eq-32}
\end{equation}
Note that the function {$g_{t}(\lambda )=\int _{0}^{\infty }\rme^{\lambda x}\, f_{t}(t)\,\rmd{}t$}
cannot be calculated in terms of elementary functions. 
We can find approximate but close to accurate result by replacing the precise 
expression~(\ref{eq-31}) by some approximation similarly to~\cite{ref-33}
\begin{equation*}
f_{t}(t)\to\frac{v_{T} }{r}\tilde{f}_{t}\left(t\frac{v_{T}}{r}\right)\,,
\end{equation*}
\begin{equation}
\tilde{f}_{t}(x)=\frac{2}{\pi ^{3/2}}\rme^{-bx}
+\frac{16(1-\rme^{-ax})^{3} }{3\pi^{3/2}x^{2}} +Ax\rme^{-cx}\,.
\label{eq-33}
\end{equation}
We choose the values of coefficients 
$A=0.70839$, $a=0.49$, $b=2.286$ and $c=1.272$. 
These coefficients ensure (1)~the same asymptotical behaviour of~$f_{t} (t)$ 
and~$\tilde{f}_{t}(tv_{T}/r)v_{T}/r$ at
infinity, (2)~coincidence of $f_{t}(t)$ and $\tilde{f}_{t}(tv_{T}/r)v_{T} /r$ 
near the zero to the second order with respect to~$t$ and (3)~correct normalization 
$\int _{0}^{\infty }\tilde{f}_{t} (x)\, \rmd{}x =1$. The plots
of the functions $r\cdot f_{t} (x\cdot r/v_{T} )/v_{T} $ and $\tilde{f}_{t}(x)$ 
are presented in \fref{fig-3}.

%%%%%%%%%%%%%%%%%%%%%%%%%%%%%%
%%%%%%%%%%%%%%%%%%%%%%%%%%%%%%

The approximation~(\ref{eq-33}) makes it possible to find the analytical
expression for $g_{t}(\lambda )$:
\begin{equation*}
g_{t}(\lambda )=\tilde{g}_{t}(\lambda r/v_{T})\,,
\end{equation*}
\begin{eqnarray}
\tilde{g}_{t}(\Lambda)&=&\frac{2}{\pi^{3/2}}\frac{1}{b-\lambda}+\frac{A}{(c-\lambda)^{2} }
\nonumber \\
&&+\frac{16}{3\pi^{3/2}} 
\bigl[
p(-\Lambda)-3p(a-\Lambda )+3p(2a-\Lambda)-p(3a-\Lambda)
\bigr]\,,
\label{eq-34}
\end{eqnarray}
where we denoted
\begin{equation}
p(\beta )=\beta\ln\beta\,.
\label{eq-35}
\end{equation}
It is easy to find $g_{\tau }(\lambda )$ from~(\ref{eq-34}) and~(\ref{eq-32}).

%%%%%%%%%%%%%%%%%%%%%%%%%%%%%%%%%%%%%%%%%%%%%%%%%%%%%%%%%%%%%%%%%%%%%%%%%%%%%%%%%%%%%%%%%%%%%
\subsection{Averaging over $\tau'$}

\begin{figure}[b]
\begin{center}
\includegraphics[scale=1]{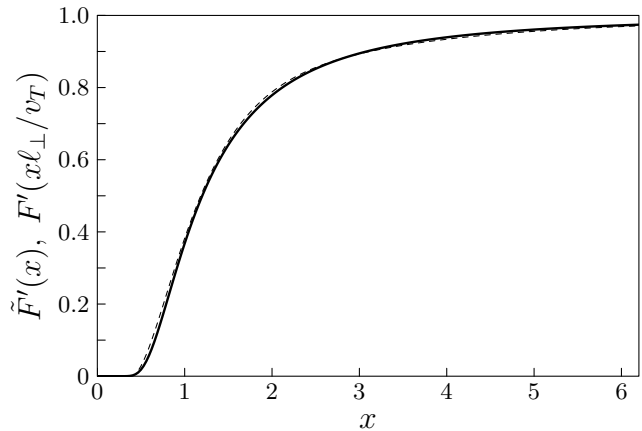}
\end{center}
\caption{Plots of functions $\tilde{F}'(x)$ (dashed line) and $F'({x\ell_{\perp}/v_{T}})$ (solid line).}
\label{fig-4}
\end{figure}

Now let us calculate the function $g_{\tau'}(\lambda)$. The time $\tau'$
{that} the atom {being} in the beam passing regime spend in the dark zone continuously is
\begin{equation}
\tau'=\frac{2}{v_{\perp}}\left(R\cos\varphi-\sqrt{r^{2}-R^{2}\sin ^{2}\varphi}\;\right)\,.
\label{eq-36}
\end{equation}
If the beam diameter is {much} smaller than the cell diameter, the expression in
brackets can be {approximately} replaces by the constant
\begin{eqnarray}
\ell_{\perp}&=&2\int _{0}^{\arcsin(r/R)}\left(R\cos\varphi-\sqrt{r^{2}-R^{2}\sin^{2}\varphi}\;\right)\Phi(\varphi)\,\rmd\varphi
\nonumber \\
&=&R\sqrt{1-\frac{r^{2}}{R^{2}}}-\frac{\pi r}{2} +\frac{R^{2}}{r} \arcsin\frac{r}{R}
\label{eq-37}
\end{eqnarray}
that equals to average transversal path length in the dark zone. This
approximation allows to readily obtain the cumulative distribution function~$F'$
of the random variable~$\tau'$:
\begin{equation}
F'(\tau ')=\rme^{-\frac{\ell _{\perp}^{2}}{\tau'^{2}v_{T}^{2}}}\,.
\label{eq-38}
\end{equation}
and
\begin{equation}
\overline{\tau}'=\frac{\ell_{\perp}}{v_{T}}\sqrt{\pi} \,.
\label{eq-39}
\end{equation}
An attempt to directly calculate the function 
$g_{\tau'}(\lambda )=\lambda\int_{0}^{\infty }(1-F'(x))\rme^{\lambda x}\,\rmd{}x$ 
does not yield the result that can be expressed through elementary functions. 
As in previous case, we can find approximate but close to accurate result by replacing 
the precise expression~(\ref{eq-38}) by
\begin{equation*}
F'(\tau ')\to\tilde{F}'\left(\tau'\frac{v_{T}}{\ell_{\perp}}\right)\,,
\end{equation*}
\begin{equation}
\tilde{F}'(x)=
\left\{
\begin{array}{ll}
1-\frac{\bigl(1-\rme^{-a(x-x_{0})}\bigr)^{4}}{(x-x_{0})^{2}}-\rme^{-b(x-x_{0})}\bigl(1+b(x-x_{0})\bigr)\,,
& x > x_{0} \\
0\,, 
& x \le{} x_{0}
\end{array}
\right.
\label{eq-40}
\end{equation}
The values of the parameters $a=0.89$, $b=2.56$ and $x_{0}=0.3864$ 
ensure smoothness of the function~$\tilde{F}'(x)$ {anywhere}, good approximation 
of~$F'(x\ell_{\perp}/v_{T})$ 
in the region where {it} is appreciably different from zero (see \fref{fig-4}) 
and get the correct value of time~$\bar{\tau }'$.      
%%%%%%%%%%%%%%%%%%%%%%%%%%%%%%%%%%%%%%%%%%%%%%%%%%%%%
%%%%%%%%%%%%%%%%%%%%%%%%%%%%%%%%%%%%%%%%%%%%%%%%%%%%%
Using the approximation function~(\ref{eq-40}), we find
\begin{equation*}
g_{\tau'}(\lambda )\approx \tilde{g}'(\lambda\ell_{\perp}/v_{T})\,,
\end{equation*}
\begin{eqnarray}
\tilde{g}'(\Lambda)&=&\Lambda \rme^{\Lambda x_{0}} 
\left[\frac{2b-\Lambda}{(b-\Lambda)^{2}}+p(4a-\Lambda )-4p(3a-\Lambda )\right.  
\nonumber \\
&&\left. -4p(a-\Lambda )+6p(2a-\Lambda )+p(-\Lambda )+\frac{1}{\Lambda}\right]\,.
\label{eq-41}
\end{eqnarray}

%%%%%%%%%%%%%%%%%%%%%%%%%%%%%%%%%%%%%%%%%%%%%%%%%%%%%%%%%%%%%%%%%%%%%%%%%%%%%%%%%%%%%%%%%%%%%
\subsection{Averaging over $\tau_{d}$}

The last function we should find for calculating the fluorescence of the atoms
in the cell is~$g_{\tau _{d}}(\lambda)$. The exact calculation of the
probability density function $f_{d} $ of the random variable~$\tau_{d}$ is a
nontrivial problem, because in the dark regime the trajectory of the atom
consists of a random number of segments of random length; along each of them the
atom moves with a random velocity. Nevertheless, it is clear that the
probability density exponentially tends to zero when~$\tau_{d}\to 0$. The
minimal distance that an atom can pass in dark regime in the transverse
direction is~$2(R-r)$. The transverse velocity~$v_{\perp}$ of the atom with the
probability greater than 98\% is less than $2v_{T}$, so we assume that 
$f_{d}(\tau_{d})=0$ when $\tau _{d}<\tau _{0}=(R-r)/v_{T}$. If the beam diameter
is much smaller than the diameter of the cell, the atom usually returns from the
dark regime into the beam passing regime after many wall collisions. The
probability for atom to return to the beam passing regime after certain
collision does not depend on the number of previous collisions. Such processes
are usually described by an exponential probability distribution function, so for
the function~$f_{d}(\tau_{d})$ of the random variable~$\tau_{d}$ we take
\begin{equation}
f_{d}(\tau _{d})=
\left\{
\begin{array}{ll}
h\rme^{-h(\tau _{d}-\tau _{0})}\,, 
& \tau _{d}>\tau _{0} \\
0\,,
& \tau _{d}<\tau _{0}\,,
\end{array}
\right.
\label{eq-42}
\end{equation}
where $h=1/(\bar{\tau }_{d}-\tau _{0} )$, $\bar{\tau }_{d}$ is the average
time of atom's being in the dark zone. From~(\ref{eq-23}) and~(\ref{eq-42}) 
it is easy to 
get~$g_{\tau _{d}}(\lambda)=\rme^{\lambda\tau _{0}}/(1-\lambda/h)$.

The value of $\bar{\tau }_{d}$ can be found from the ``time balance''
conditions. It means that the total time that an atom spends in the beam or {in} the
dark zone is proportional to the volume of this zone. Consider the cycle in
which an atom being in the beam passing regime undergoes $N$ elastic
collisions with the wall, and then moves into the dark regime. The history of
any atom can be represented as a sequence of such cycles. The total time~$t_{b}$ 
spent by an atom in the beam zone over the cycle is $t_{b} = (N+1)\tau $. In
turn, the time~$t_{d}$ spent by the atom in a dark zone during the cycle is
$t_{d}=\tau _{d}+N\tau'$. 
The ratio of averaged times~$t_{b}$ and~$t_{d}$
over the different cycles must be equal to the ratio of volumes of the bright
and dark zones. Hence, we find
\begin{equation}
\bar{\tau}_{d}=\left(\frac{R^{2}}{r^{2}}-1\right)\bar{\tau }(\bar{N}+1)-\bar{\tau }'\bar{N} ,
\label{eq-43}
\end{equation}
where $\bar{N}=\alpha /(1-\alpha )$ is the average number of successive elastic collisions with the cell wall.

In the conclusion of this section we write down the expression for the averaged excited state population:
%%%
%
\begin{eqnarray}
\lefteqn{\langle\rho_{33}\rangle= 
\int\limits_{-\infty}^{\infty} M_1(v_z) \, \Hat{U}^T \Ghathat_t (1-\alpha)
\bigg(\Ihathat-\alpha \, \Ghathat'_{\tau'}\Ghathat'_{\tau'}\bigg)^{-1} dv_z \cdot \ \langle \hat{\rho}_b \rangle} \nonumber \\
&&+ \int\limits_{-\infty}^{\infty} M_1(v_z)  V dv_z+
\int\limits_{-\infty}^{\infty} M_1(v_z) \, \Hat{U}^T\Bigg[\left(\Ihathat-\Ghathat_t
\right)+\Ghathat_{t} \Ghathat'_{\tau'} \label{eq-15-new}\\
&&\times \left(\Ihathat-\Ghathat_{\tau} \Ghathat'_{\tau'} \right)^{-1} 
\Big[ \Ihathat-(1-\alpha)\bigg(\Ihathat-\alpha \Ghathat_{\tau} \Ghathat'_{\tau'} \bigg)^{-1} \Big]  \Bigg] \Hat{\rho}_s \, dv_z, \nonumber
\end{eqnarray}
%
%%%
where $M_1(v_z)=\rme^{-v_z/v_T}/(v_T\sqrt{\pi})$ is one-dimensional Maxwell distribution function,
%%%
%
\begin{eqnarray}
\Ghathat_t&=&
\Xhathat
\left(
\begin{array}{ccc}
g_t(\lambda_1) & 0 & 0   \\
0 & g_t(\lambda_2) & 0 \\
0 & 0 & g_t(\lambda_3)
\end{array}
\right)
\Xhathat^{-1} ,  \\
\Ghathat_\tau&=&
\Xhathat
\left(
\begin{array}{ccc}
g_\tau(\lambda_1) & 0 & 0   \\
0 & g_\tau(\lambda_2) & 0 \\
0 & 0 & g_\tau(\lambda_3)
\end{array}
\right)
\Xhathat^{-1}
% \label{eq-22-new} 
,  \\
\Ghathat'_{\tau'}&=&
\Xhathat'
\left(
\begin{array}{ccc}
g_{\tau'}(\lambda'_1) & 0 & 0   \\
0 & g_{\tau'}(\lambda'_2) & 0 \\
0 & 0 & g_{\tau'}(\lambda'_3)
\end{array}
\right)
\Xhathat'^{-1}
% \label{eq-22-new} , 
\\
\langle \Hat{\rho}_b \rangle &=&\bigg(\Ihathat-\langle\Ghathat'_{\tau_d}\rangle \Dhathat\bigg)^{-1} \langle\Ghathat'_{\tau_d}\rangle \Hat{C},
\end{eqnarray}
where expressions (\ref{eq-19},\ref{eq-20}) for $\Hat{C}$ and $\Dhathat$ can be rewritten as
\begin{eqnarray}
\Hat{C}&=&\int\limits_{-\infty}^{\infty}M_1(v_z) \Big(\Ihathat-\Ghathat_{\tau}\Ghathat'_{\tau'} \Big)^{-1}
\Big[\Ihathat-(1-\alpha)\Ghathat_{\tau}\Ghathat'_{\tau'} \nonumber \\
 && \times \Big( \Ihathat-\alpha \Ghathat_\tau\Ghathat'_{\tau'} \Big)^{-1} \Big]\big( \Ihathat-\Ghathat_{\tau}\big)\hat{\rho}_s \; dv_z, \\
\Dhathat&=&\int\limits_{-\infty}^{\infty}M_1(v_z) \; \Ghathat_{\tau} (1-\alpha) \Big(\Ihathat-\alpha \Ghathat'_{\tau'} \Ghathat_\tau \Big)^{-1} \, dv_z, \\
\langle\Ghathat'_{\tau_d}\rangle&=&\int\limits_{-\infty}^{\infty} M_1(v_z)
\Xhathat'
\left(
\begin{array}{ccc}
g_{\tau_d}(\lambda'_1) & 0 & 0   \\
0 & g_{\tau_d}(\lambda'_2) & 0 \\
0 & 0 & g_{\tau_d}(\lambda'_3)
\end{array}
\right)
\Xhathat'^{-1} \, dv_z.
\end{eqnarray}

%%%%%%%%%%%%%%%%%%%%%%%%%%%%%%%%%%%%%%%%%%%%%%%%%%%%%%%%%%%%%%%%%%%%%%%%%%%%%%%%%%%%%%%%%%%%%
%%%%%%%%%%%%%%%%%%%%%%%%%%%%%%%%%%%%%%%%%%%%%%%%%%%%%%%%%%%%%%%%%%%%%%%%%%%%%%%%%%%%%%%%%%%%%
\section{Results of calculations.}
\label{sect-5}

An absorption of the laser power in the cell filled by the gas of three-level
$\Lambda$-atoms is proportional to the average excited state population
described by expression~(\ref{eq-15}). Typical parameters of $^{87}$Rb
atom in vacuum cell were taken as the parameters of the $\Lambda$-atom in our calculations. The
atomic mass~$m$ was assumed to be equal to the mass of $^{87}\mathrm{Rb}$ isotope, the
temperature $T=20^{\circ}\mathrm{C}$, the ground state relaxation rate $\Gamma=300\,\mathrm{s}^{-1}$, the
optical coherence relaxation rate $\gamma '=1.8\times 10^{7}\,\mathrm{s}^{-1}$. We assume
$V_{1}=V_{2}=V$ and we set the optical detuning $\Omega _{L}=0$. Calculations
were performed for the cell radius~$R=0.5$~cm and for three values of the laser
beam radius $r$ ($r=0.5$, $1.5$ and $5\,\mbox{mm}$) and for five values of the probability
$\alpha$ of the atomic-wall elastic collision ($\alpha=0$, $0.25$, $0.5$, $0.75$
and~$1$).

\begin{figure}
\begin{center}
\begin{tabular}{cc}
\includegraphics[scale=0.7]{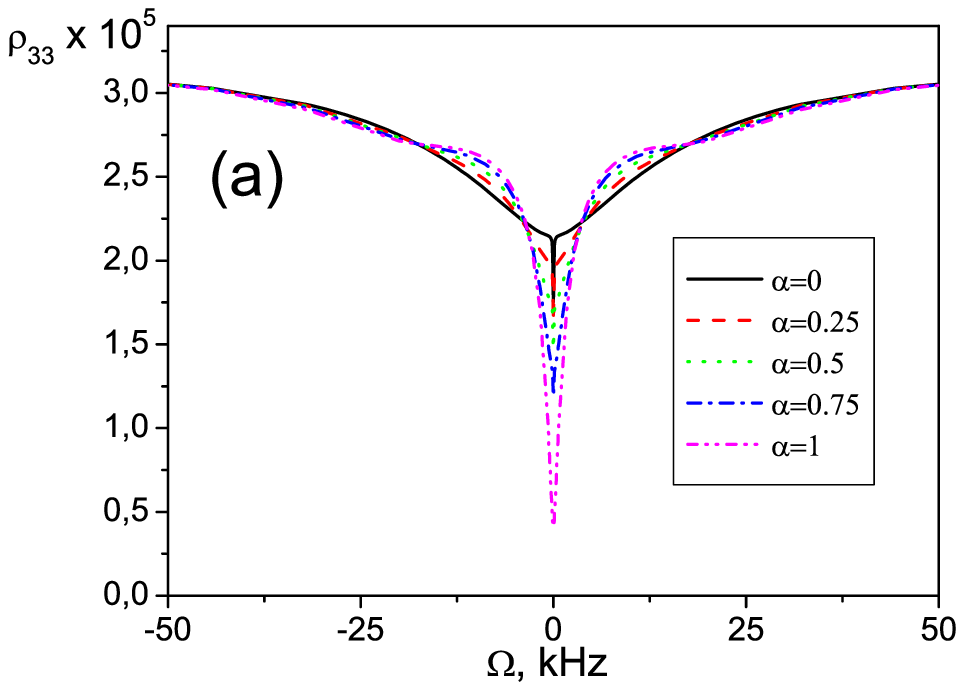} &
\includegraphics[scale=0.7]{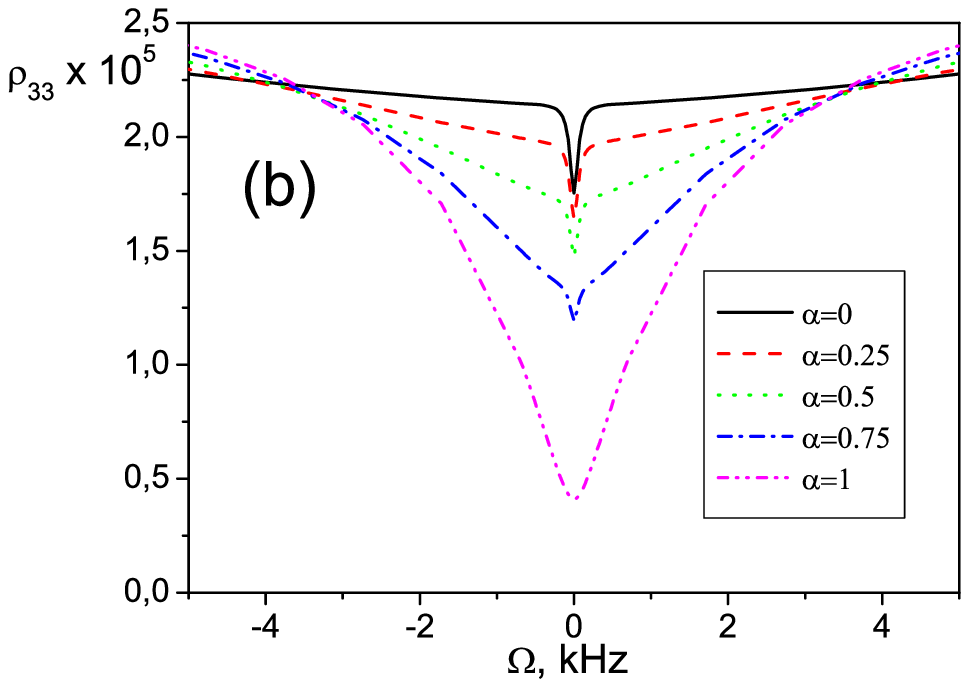} \\
\includegraphics[scale=0.7]{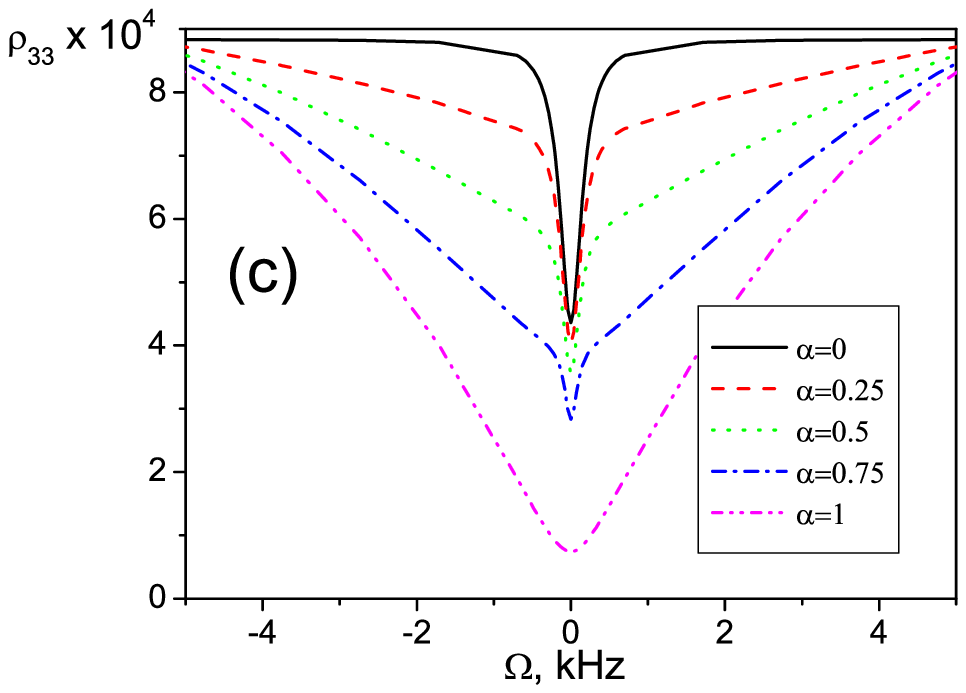} &
\includegraphics[scale=0.7]{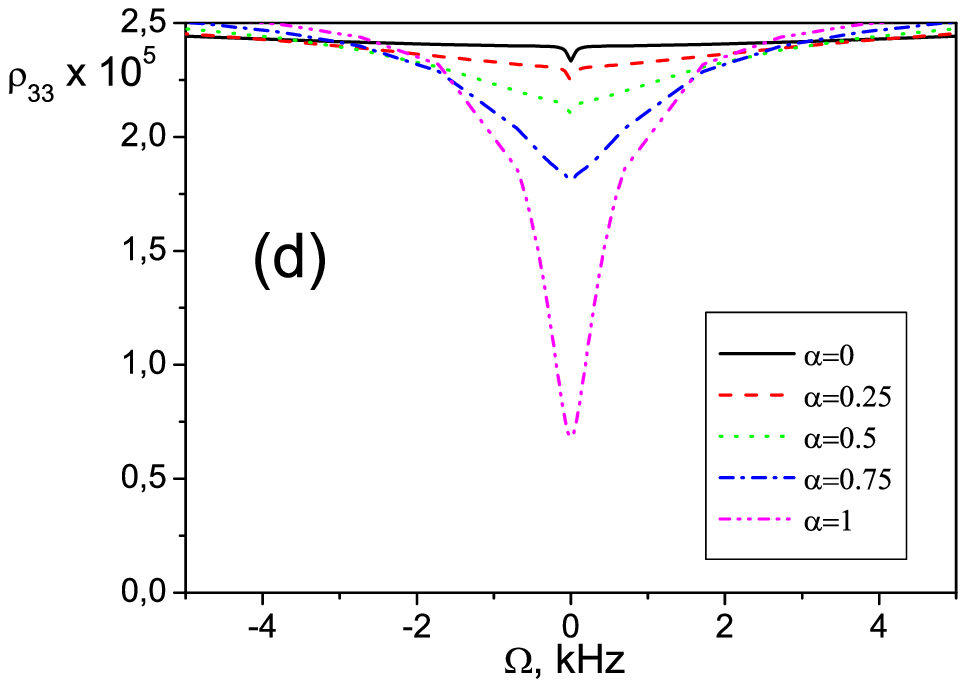} \\
\includegraphics[scale=0.7]{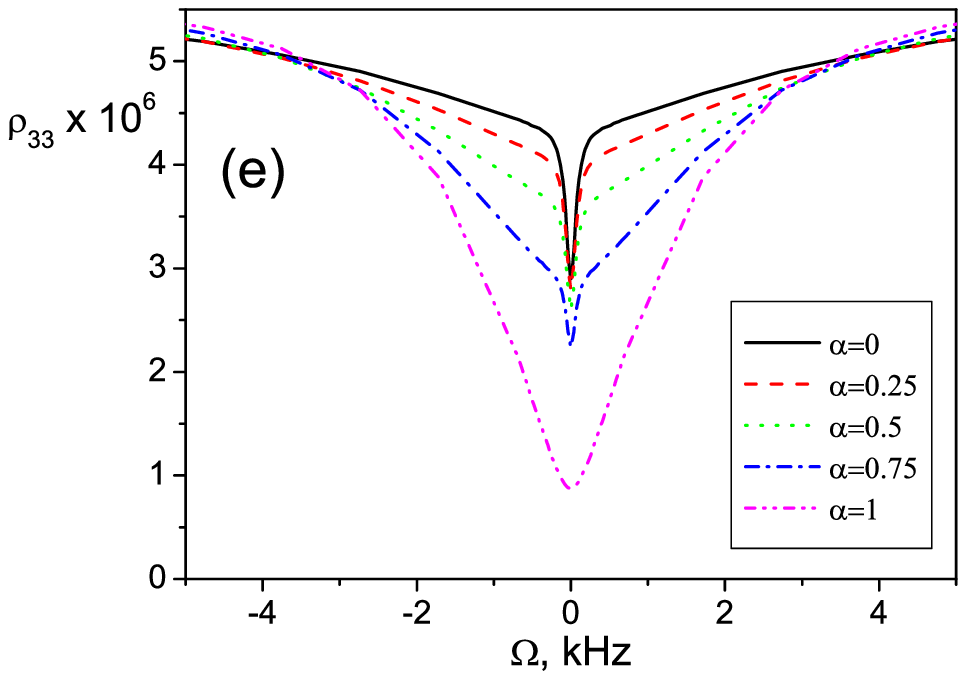} &
\includegraphics[scale=0.7]{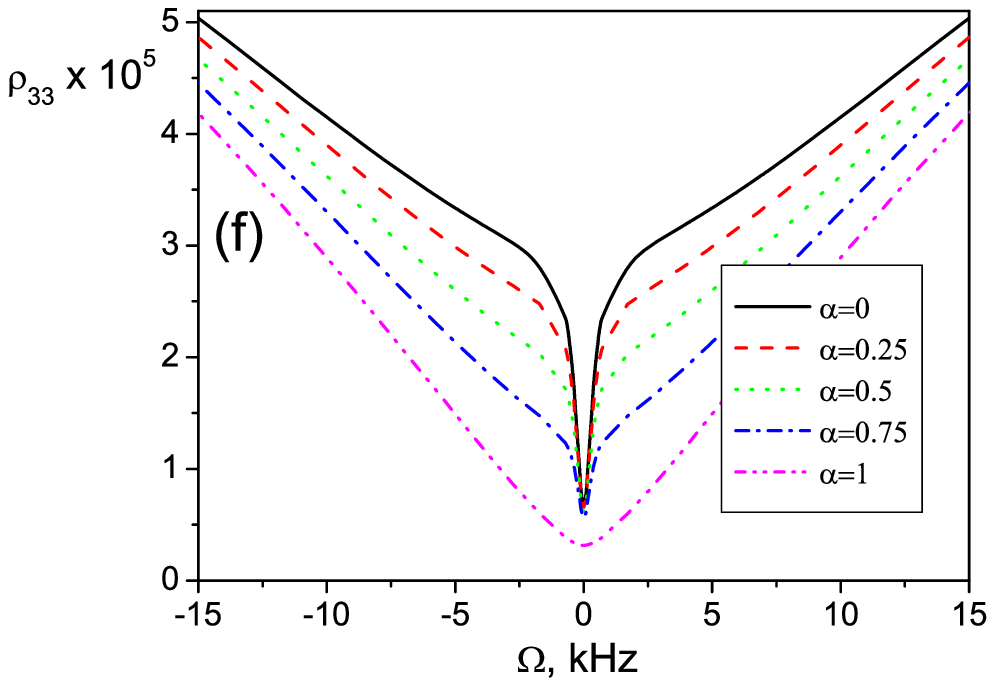}
\end{tabular}
\end{center}
\caption{The shape of CPT resonance for different $\alpha$, $V$ and $r$. 
Parameters: $r=1.5\,\mbox{mm}$, $V=7.7\times 10^{5}\,\mbox{s}^{-1} $ 
(a, b); $r=0.5\, \mbox{mm}$, $V=1.46\times 10^{7}\,\mbox{s}^{-1}$ 
(c); $r=0.5\, \mbox{mm}$, $V=7.3\times 10^{5}\, \mbox{s}^{-1}$ 
(d); $r=5 \,\mbox{mm}$, $V=3.3\times 10^{5}\, \mbox{s}^{-1}$ 
(e); $r=5\, \mbox{mm}$, $V=1.46\times 10^{6}\, \mbox{s}^{-1}$~(f);}
\label{fig-5}
\end{figure}

The structure of the CPT resonance for $\alpha =0$ was discussed in~\cite{ref-11,ref-12,ref-13,ref-14}.
Resonance consists of a broad pedestal whose width is about several tens of
kilohertz and of a narrow central peak (see \fref{fig-5}(a)). 
%%%%%%%%%%%%%%%%%%%%%%
%%%%%%%%%%%%%%%%%%%%%%
Broad background is formed
by the atoms that have passed through the beam zone only once, whereas the
central peak is formed by multiple passes. {If} the elastic collision probability~$\alpha$ 
is nonzero, the additional ``intermediate'' structure in CPT resonance
appears. This is a peak whose width is greater than the width of the narrow
central peak, but less than the width of the {broad background. This
peak is formed by atoms that pass the beam zone  many times in beam passing
regime.} It should be noted that in the elastic collision act the atomic longitudinal
velocity~$v_{z}$ remains constant, in contrast with the case of inelastic
scattering when atoms pass the beam zone just once in the beam passing regime.
This fact explains the occurrence of ``intermediate'' structure depending on~$\alpha$ 
when the cell radius coincides with the radius of the beam (see \fref{fig-5}(e) and (f)). 
Indeed, in the considered case of the laser field with narrow spectral linewidth only
the atoms with small $|v_z|$ contribute to the CPT resonance.
The ``intermediate'' structure is formed by the atoms that remains in this velocity group
during several passage through the cell volume whereas the narrow structure is formed by 
the atoms that change their velocities after one wall collision.

At Fig. 5(a)--(f) we see that the higher elastic collision probability $\alpha$
corresponds to a lower excited state population at $\Omega=0$. To explain it we should note that for 
higher $\alpha$ atoms spent longer time in the beam passing regime; therefore, they have more time 
to came into the dark state.

It should be noted that if $\alpha=1$, the narrow peak disappears. It can be
explained by the fact that all the atoms contributing to the CPT resonance
formation are in the beam passing regime permanently. These atoms spent relative
higher part of time in the beam zone than all the atoms in case of~$\alpha=0$.
Therefore, the narrow peak is formed by atoms {which} passed the beam, then {spend}
quite {a} long time in the dark regime and then returned back to the beam.
Intermediate structure is formed by atoms which passed the beam {several times} repeatedly in
the beam passing regime. The time between two subsequent beam passage occurs to
be smaller than in the case of inelastic collision. Therefore, the resonance
occurs to be broader. The broad pedestal is formed by atoms {that} pass the beam
zone once during the coherence time.

%%%%%%%%%%%%%%%%%%%%%%%%%%%%%%%%%%%%%%%%%%%%%%%%%%%%%%%%%%%%%%%%%%%%%%%%%%%%%%%%%%%%%%%%%%%%%
%%%%%%%%%%%%%%%%%%%%%%%%%%%%%%%%%%%%%%%%%%%%%%%%%%%%%%%%%%%%%%%%%%%%%%%%%%%%%%%%%%%%%%%%%%%%%
\section{Conclusions}
\label{sect-6}

We developed a theory of {the} coherent population trapping resonance in a cylindrical
cell filled with antirelaxation coated walls. The laser beam is assumed to have
constant intensity in cylindrical region. The atom-wall interaction is described
by the probability of elastic collisions. It is shown that the nonzero
probability of elastic collision leads to distortion of the CPT resonance
lines that consists in appearance of the additional peak formed by the atoms,
repeatedly passing through the beam zone in the beam passing regime. The general
scheme of calculation can be applied not only to CPT formation in wall-coated
cells but also for the cell with small buffer gas pressure (much lower than the
necessary for the diffusion approximation applicability). Another possible
application can be connected with the high-precision clock {based} on {an ensemble} of cold
atoms or ions, e.g. ions of $^{229}\mbox{Th}$~\cite{ref-34}. The energy splitting of the 
$^{229}\mbox{Th}$
ground-state doublet is about $7.6\mbox{eV}$~\cite{ref-35} corresponding to vacuum ultraviolet
and long, about an hour, expected half-life of the excited state of the nuclei~\cite{ref-36} 
makes this nuclei a possible promising object for nuclear frequency
standard. The study of one-photon resonances in the two-dimensional trap with
zonal pumping may be applicable to the spectroscopy of this nucleus.

%%%%%%%%%%%%%%%%%%%%%%%%%%%%%%%%%%%%%%%%%%%%%%%%%%%%%%%%%%%%%%%%%%%%%%%%%%%%%%%%%%%%%%%%%%%%%
\ack

This work was supported by the grant of the Russian Federation for young
candidates of science MK-5318.2010.2, the Federal Program ``Scientific and
scientific-pedagogical personnel of innovative Russia in 2009-2013'' and the
grant of Russian Foundation for Basic Research 11-02-90426\_Ukr\_f\_a, NASU
projects V136, VC139, State Foundation of Fundamental Researches of Ukraine
(project F40.2/039).

\section*{References}

%%%%%%%%%%%%%%%%%%%%%%%%%%%%%%%%%%%%%%%%%%%%%%%%%%%%%%%%%%%%%%%%%%%%%%%%%%%%%%%%%%%%%%%%%%%%%

\end{document}